\documentclass[aps,prd,nofootinbib,twocolumn,superscriptaddress,preprintnumbers,letterpaper,natbib,nofootinbib]{revtex4-1}
\pdfoutput = 1
\usepackage{amssymb}
\usepackage{amsmath}
\usepackage[usenames,dvipsnames]{color}
\usepackage{hyperref}
\usepackage{graphicx}

\usepackage[page,toc,titletoc,title]{appendix}

\hypersetup{
    colorlinks=true,       
    linkcolor=Cyan,          
    citecolor=Magenta}   
    
\makeatletter

\newcommand{\bea}{\begin{eqnarray}}
\newcommand{\eea}{\end{eqnarray}}

\begin{document}

\title{Neff in the Standard Model at NLO is 3.043}

\author{Mattia Cielo}
\email{mattia.cielo@unina.it}
\thanks{ORCID: \href{https://orcid.org/0000-0002-5036-3550}{0000-0002-5036-3550}}
\affiliation{Dipartimento di Fisica E. Pancini, Naples, and INFN, Sezione di Napoli, Monte S.Angelo I-80126 Naples, Italy}

\author{Miguel Escudero}
\email{miguel.escudero@cern.ch}
\thanks{ORCID: \href{https://orcid.org/0000-0002-4487-8742}{0000-0002-4487-8742}}
\affiliation{Theoretical Physics Department, CERN, 1211 Geneva 23, Switzerland}

\author{Gianpiero Mangano}
\email{gmangano@na.infn.it}
\thanks{ORCID: \href{https://orcid.org/0000-0002-6901-4633}{0000-0002-6901-4633}}
\affiliation{Dipartimento di Fisica E. Pancini, Naples, and INFN, Sezione di Napoli, Monte S.Angelo I-80126 Naples, Italy}

\author{Ofelia Pisanti}
\email{pisanti@na.infn.it}
\thanks{ORCID: \href{https://orcid.org/0000-0002-2745-9204}{0000-0002-2745-9204}}
\affiliation{Dipartimento di Fisica E. Pancini, Naples, and INFN, Sezione di Napoli, Monte S.Angelo I-80126 Naples, Italy}

\begin{abstract} 
\noindent The effective number of relativistic neutrino species is a fundamental probe of the early Universe and its measurement represents a key constraint on many scenarios beyond the Standard Model of Particle Physics. In light of this, an accurate prediction of $N_{\rm eff}$ in the Standard Model is of pivotal importance. In this work, we consider the last ingredient needed to accurately calculate $N_{\rm eff}^{\rm SM}$: standard zero and finite temperature QED corrections to $e^+e^- \leftrightarrow  \nu\bar{\nu}$ interaction rates during neutrino decoupling at temperatures around $T\sim {\rm MeV}$. We find that this effect leads to a reduction of $-0.0007$ in $N_{\rm eff}^{\rm SM}$. This NLO QED correction to the interaction rates, together with finite temperature QED corrections to the electromagnetic density of the plasma, and the effect of neutrino oscillations, implies that $N_{\rm eff}^{\rm SM} = 3.043$ with a theoretical uncertainty that is much smaller than any projected observational sensitivity.  
\end{abstract}

\preprint{CERN-TH-2023-103}

\maketitle

{
  \hypersetup{linkcolor=black}
}

\setlength\parskip{4pt}

\section{Introduction}
The number of effective relativistic neutrino species, $N_{\rm eff}$, represents a key probe of the thermal history of the early Universe. In particular, an array of new light states beyond the Standard Model (BSM) are expected to contribute to this quantity and current measurements of $N_{\rm eff}$ at recombination and during big bang nucleosynthesis (BBN) represent in many cases our best handle on many BSM settings~\cite{Allahverdi:2020bys}. Relevant examples of these include light sterile neutrinos~\cite{Barbieri:1990vx,Gariazzo:2019gyi}, dark sectors~\cite{Berezhiani:2000gw,Brust:2013ova}, and pseudo-Goldstone bosons including axions and majorons~\cite{Weinberg:2013kea,DEramo:2021psx,Notari:2022ffe,DiLuzio:2022gsc,DEramo:2022nvb,Escudero:2019gfk}. Yet, $N_{\rm eff}$ measurements are currently the best test of high frequency primordial gravitational wave backgrounds~\cite{Caprini:2018mtu}, they provide a lower bound on the reheating temperature of the Universe~\cite{Giudice:2000ex,deSalas:2015glj,Hasegawa:2019jsa}, or they tell us how light can thermal dark matter be~\cite{Kolb:1986nf,Sabti:2019mhn}.

From the observational perspective, the Planck Collaboration has reported unprecedented precision measurements of $N_{\rm eff}$. Within the framework of the standard cosmological model, $\Lambda$CDM: $N_{\rm eff} = 2.99 \pm 0.17 $ at 68\% CL~\cite{Planck:2018vyg}. Furthermore, other recent ground-based CMB experiments such as SPT-3G~\cite{SPT-3G:2021wgf} and ACT~\cite{ACT:2020gnv} report measurements of $N_{\rm eff}$ compatible with Planck albeit with error bars which are a factor of $\sim 2-3$ larger. The situation is expected to improve very soon as the Simons Observatory is under construction and aiming to deliver a measurement of $N_{\rm eff}$ with a precision of $\sigma(N_{\rm eff}) \simeq 0.05$ in the upcoming $\sim 6$ years~\cite{SimonsObservatory:2018koc}. Importantly, the next generation of CMB experiments should reach a precision of $\sigma(N_{\rm eff}) \simeq 0.03$~\cite{CMB-S4:2016ple,Abazajian:2019eic}. Looking further ahead, ultrasensitive futuristic CMB experiments could even reach a sensitivity of $\sigma(N_{\rm eff}) \simeq 0.014$~\cite{Sehgal:2020yja}.

In order to draw meaningful conclusions from precise measurements of $N_{\rm eff}$ its value in the Standard Model needs to be known accurately. In particular, $N_{\rm eff}^{\rm SM}$ as relevant for CMB observations is defined as:
\begin{align}\label{eq:Neff}
    N_{\rm eff}^{\rm SM} \equiv \frac{8}{7}\left(\frac{11}{4}\right)^{4/3} \left(\frac{\rho_\nu}{\rho_\gamma}\right) \,,
\end{align}
where $\rho_\nu$ and $\rho_\gamma$ are the energy densities in relativistic neutrinos and photons at $T \ll m_e$, respectively. 

A precision calculation of $N_{\rm eff}^{\rm SM}$ requires one to solve the process of neutrino decoupling in the early Universe and this has been a subject of intense study for more than 40 years now, see~\cite{Dicus:1982bz,Rana:1991xk,Dodelson:1992km,Dolgov:1992qg,Hannestad:1995rs,Dolgov:1997mb,Dolgov:1998sf,Gnedin:1997vn,Esposito:2000hi,Mangano:2001iu,Mangano:2005cc,Birrell:2014uka,Grohs:2015tfy,deSalas:2016ztq,Froustey:2019owm,Bennett:2019ewm,Escudero:2018mvt,EscuderoAbenza:2020cmq,Akita:2020szl,Froustey:2020mcq,Bennett:2020zkv} for references devoted to this problem, see~\cite{Dolgov:2002wy,Akita:2022hlx} for reviews and~\cite{Lesgourgues:2013sjj} for a book.

Neglecting the process of neutrino decoupling, one obtains $N_{\rm eff}^{\rm SM} = 3$ simply by entropy conservation between $T\sim 10\,{\rm MeV}$ and $T\ll m_e$~\cite{Kolb:1990vq,Gorbunov:2011zz}. However, a series of physical ingredients make this number larger by $\sim 1\%$ as summarized in Table~\ref{tab:DNeffeffects} and as discussed in what follows:

First, neutrino interactions freeze out at a temperature around $T\sim 2\,{\rm MeV}$, which is not too different from the electron mass. This means that some electrons and positrons do annihilate to neutrinos and slightly heat the neutrino bath, increasing their energy density and therefore making $N_{\rm eff}$ larger; see Eq.~\eqref{eq:Neff}. This is by far the largest effect and leads to $\Delta N_{\rm eff}^{\rm SM} = +0.03$~\cite{Dicus:1982bz,Rana:1991xk,Dodelson:1992km,Hannestad:1995rs,Dolgov:1997mb,Dolgov:1998sf,Escudero:2018mvt,EscuderoAbenza:2020cmq}.

Second, $N_{\rm eff}$ is a ratio between neutrino and electromagnetic energy densities. At the time of neutrino decoupling the electromagnetic sector of the plasma consist of a large number of photons, electrons, and positrons, and at finite temperature these particles obtain small but nonzero corrections to their dispersion relations and to their masses, e.g. $m_\gamma(T_\gamma) \simeq e  T_\gamma/\sqrt{6} \simeq 0.12\,T_\gamma$ where $e\simeq 0.3$ is the electric charge coupling. This mass is substantially smaller than the typical energy of the particles $E\sim 3 T$ but nevertheless can change the pressure and energy density by $\sim 1/30$ and this indeed leads to a positive correction to $N_{\rm eff}^{\rm SM}$ of +0.01~\cite{Heckler:1994tv,Bennett:2019ewm}.

Third, taking the measured values of the masses and mixings of neutrinos one readily realizes that neutrinos start oscillating at $T\sim 5-10\,{\rm MeV}$. This means that neutrino oscillations are also a relevant physical ingredient at the time of neutrino decoupling. The numerical impact of neutrino oscillations on $N_{\rm eff}^{\rm SM}$ is not large as what neutrino oscillations do is simply redistribute energy among neutrino flavors. When this effect is included it leads to a positive correction to $N_{\rm eff}^{\rm SM}$ of $\sim +0.0007$~\cite{Mangano:2001iu,Mangano:2005cc,deSalas:2016ztq,Akita:2020szl,Froustey:2020mcq,Bennett:2020zkv}. 

Finally, the key processes that control neutrino decoupling are annihilations and scatterings with electrons and positrons. So far, all calculations of $N_{\rm eff}$ in the Standard Model have used the Born rates as computed using Fermi's theory of weak interactions at leading order. However, and as is well known for neutrino experiments sensitive to neutrino-electron scatterings~\cite{Super-Kamiokande:1998zvz,LSND:2001akn,SNO:2008gqy,Bellini:2011rx}, there are relevant QED corrections to these types of processes which can alter the interaction rates by $\sim 5\%$ for MeV energies; see e.g.~\cite{Bahcall:1995mm,Passera:2000ug}. The potential relevance of these radiative corrections was highlighted in~\cite{EscuderoAbenza:2020cmq} but to date, there is no study accounting for these QED corrections when solving the process of neutrino decoupling. This is precisely the gap that we fill in this paper. We follow closely the calculation of radiative corrections to $e^+e^- \to \nu\bar{\nu}$ in a stellar plasma~\cite{Esposito:2003wv} to find the relevant NLO correction to the $e^+e^-$ reaction rates. At next-to-leading order (NLO) the $e^+e^-\to \nu\bar{\nu}\gamma$ phase-space integration is extremely challenging and impedes the use of the Liouville equation to solve for the process of neutrino decoupling. In this situation, we instead solve the process of neutrino decoupling assuming that neutrinos possess a thermal distribution function with a dynamical temperature driven by neutrino interactions as considered in~\cite{Escudero:2018mvt,EscuderoAbenza:2020cmq}. The method has been shown to accurately reproduce the results obtained by solving the Liouville equation and remarkably allows us to implement the NLO effects in a straightforward manner. Our final result shows that the effect of this NLO correction on the rates is to reduce $N_{\rm eff}^{\rm SM}$ by $-0.0007$, which is similar in magnitude to the effect of neutrino oscillations. Putting all effects together, we predict 
\begin{align}
    N_{\rm eff}^{\rm SM} = 3.0432(2)\simeq 3.043\,,
\end{align}
where the reminder theoretical uncertainties are expected at the level of $0.0002$ arising mainly from a combination of a lack of precise knowledge in $\theta_{12}$~\cite{Bennett:2020zkv} and yet unexplored QED corrections to $e\nu \to e\nu$ reaction rates. 

The reminder of our paper is as follows. In Section~\ref{sec:NLOrate}, we describe the calculation of the $e^+e^-\to \nu\bar{\nu}(\gamma)$ reaction rate at NLO in the fine structure constant $\alpha$. Then in Section~\ref{sec:Neffresults}, we describe our modelling of neutrino decoupling and implement this new NLO rate. We draw our conclusions in Section~\ref{sec:conclusions}.

\begin{table}[t]
    \centering
    \begin{tabular}{c|c} \hline  \hline
        Physical Scenario     & $\,\,\,\, N_{\rm eff}^{\rm SM}-3\,\,\,\, $  \\ \hline \hline
        Instantaneous neutrino decoupling & $0$  \\ \hline 
        Neutrino interactions : residual $e^+e^-\to \bar{\nu}\nu$ & $0.03$  \\ \hline 
        QED corrections to $\rho_{e^+e^-}$ and $\rho_\gamma$ & $0.01$  \\ \hline
        Neutrino oscillations  & $\,\,0.0007$  \\ \hline
        QED corrections to $e^+e^-\leftrightarrow \bar{\nu}\nu$ rates (this work) & $\!\!-0.0007$  \\ \hline \hline
    \end{tabular}
    \caption{Various contributions to $N_{\rm eff}$ in the Standard Model. We highlight each effect to the first digit. When all are taken together we find $N_{\rm eff}^{\rm SM} = 3.043$.}
    \label{tab:DNeffeffects}
\end{table}

\section{Methodology and Calculation of $e^+e^-\to \nu \bar{\nu}$ at NLO}\label{sec:NLOrate}

In this section, we briefly summarize the results of \cite{Esposito:2003wv,Esposito:2001if} and outline their extension to the parameter space needed for the present calculation. The energy transfer rate from electrons to neutrinos by $e^+e^- \to \bar{\nu} \nu$ annihilations is defined as: 
\bea Q && \equiv \left.\frac{\delta \rho}{\delta t}\right|_{e^+e^- \to \bar{\nu}\nu}  \nonumber \\
&& = \int \frac{d^3 {\mathbf{p}}_1}{(2 \pi)^3 2 E_1} \int \frac{d^3 {\mathbf{p}}_2}{(2 \pi)^3 2 E_2}
\,(E_1 + E_2) \, \mathcal{F}_{e^-}(E_1) \, \mathcal{F}_{e^+}(E_2)
\nonumber \\
&& \frac{1}{(2 \pi)^2} \int \frac{d^3{\mathbf{q}}_1}{2\,\omega_1} \int \frac{d^3{\mathbf{q}}_2}{2\,\omega_2} \,
\delta^{(4)}(p_1+p_2-q_1-q_2) \,
\nonumber \\
&& (1-\, \mathcal{F}_\nu(q_1)) \, (1-\mathcal{F}_{\bar{\nu}}(q_2))\, \sum_{{\mathrm spin,\nu_\alpha}} {|M|}^2
\label{eq:Qloss}
\eea
where here $\mathcal{F}$ are distribution functions and $|M|^2$ is the matrix element squared.

\begin{figure}[t]
\centering
\includegraphics[width=0.48\textwidth]{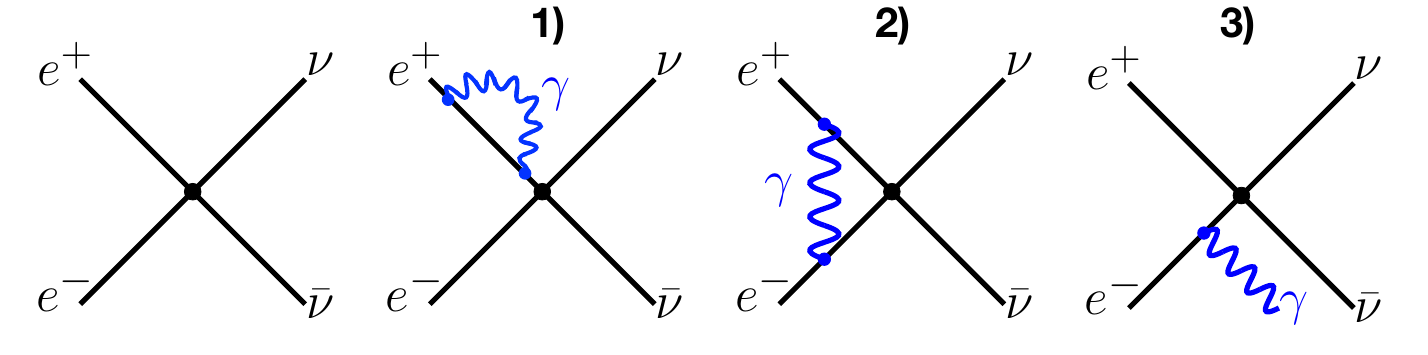}
\vspace{-0.4cm}
\caption{
Born diagram for $e^+e^-\to \bar{\nu} \nu$ annihilations and NLO QED corrections to the $e^+e^-\to \bar{\nu} \nu$ process, including processes contributing to: 1) electron mass and wave function renormalization, 2) electromagnetic vertex correction, and 3) photon emission and absorption. 
}
\label{fig:cartoon}
\end{figure}

 NLO radiative corrections to the energy loss rate, $Q$, induced by the process $e^+e^- \to \nu\bar{\nu}$ can be classified as follows: 1) electron mass and wave function renormalization, 2) electromagnetic vertex correction, and 3) $\gamma$ emission and absorption. These corrections are illustrated in Fig.~\ref{fig:cartoon}. In~\cite{Esposito:2003wv}, the real time formalism for finite temperature quantum field theory was used, consisting of adding to the standard Feynmann propagators (which produce the zero-temperature or vacuum radiative corrections) an additional contribution, depending on temperature, that describes the interactions with the real particles of the thermal bath (finite-temperature radiative corrections).

All the amplitudes involved were evaluated analytically, as described in detail in \cite{Esposito:2003wv}, and their square was then integrated by the Monte Carlo technique over the relevant phase space with an accuracy of better than 1\% on  $Q$ \cite{Esposito:2001if}. It is worth noting that each correction in the list above is plagued by infrared divergences, even if the final result is of course divergence free. To deal with these divergences, they were regularized by explicitly subtracting all divergent terms in a Laurent series around the pole singularities in contributions of the type (1) and (2) at finite temperature, while combining the soft part of bremsstrahlung (SB) with the vacuum radiative corrections (VC) in order to estimate the infrared-free zero-temperature radiative corrections. A check was made that the final result is independent of the soft cutoff, introduced in the numerical evaluation of the SB+VC contribution, as long as it is much smaller than the electron mass \cite{Passera:2000ug}.

The relevant temperatures to be investigated for the present analysis were not completely covered in \cite{Esposito:2003wv,Esposito:2001if} (and was partially represented by the top-left corner in Fig.~7 of \cite{Esposito:2003wv}), while the values of the electron chemical potential of interest in the early Universe are the lowest one of the previous work.  We have thus extended our previous calculation to the relevant region, up to plasma temperatures $T \sim 20 \,{\rm MeV}$, and show the resulting relative correction $\xi^{\rm NLO} = (Q|_{\rm NLO} -Q|_{\rm LO})/Q|_{\rm LO}$ in Fig.~\ref{fig:NLOrate}. We can see that the correction can be as large as $-5$\% for $T_\gamma \sim 2$\,{\rm MeV} as relevant for neutrino decoupling. In particular, the bulk of these NLO corrections comes from zero temperature and  bremsstrahlung contributions ($\sim 80\%$ for $T_\gamma \sim 2$\,{\rm MeV}), while finite temperature radiative corrections remain subdominant. Since the NLO rate is smaller than the LO one, we would then expect these radiative corrections to lower the value of $N_{\rm eff}$. In addition, note that in principle the QED corrections to the energy exchange rate are neutrino flavor dependent, as the weak vertex has different coupling for $\nu_e$ and $\nu_{\mu, \tau}$. However, in practice, the numerical difference between neutrino flavors turns out to be substantially smaller than $1\%$ for the relevant temperature range, and therefore the NLO correction depicted in Fig.~\ref{fig:NLOrate} effectively applies to all neutrino flavors. 

Note that the analytical calculation carried out in \cite{Esposito:2003wv} was performed under the assumption that in Eq.~(\ref{eq:Qloss}), $\mathcal{F}_\nu(q_1)=\mathcal{F}_{\bar{\nu}}(q_2)=0$; i.e. we can neglect neutrino Pauli blocking effects. This was justified by observing that the neutrino mean free path in stars is large enough so they do not experience any further interaction after emission, and thus, their number density at emission is negligible. This is, however, not the case in the early Universe, where a thermal population of neutrinos is present. Adding the two factors of $1-\mathcal{F}_\nu$ to the calculation is extremely challenging, as this would require enhancing the numerical integration by two dimensions. The total effect can be as large as $\sim 10\%$ in the total rate. However, since what is relevant is the ratio NLO vs leading order (LO), the impact of neglecting Pauli blocking on the ratio is negligible and should contribute to at most a 1\% uncertainty on the ratio shown in Fig.~\ref{fig:NLOrate}. 

\section{Neutrino Decoupling at NLO}\label{sec:Neffresults}

\begin{figure}[t]
\centering
\hspace{-0.5cm} \includegraphics[width=0.49\textwidth]{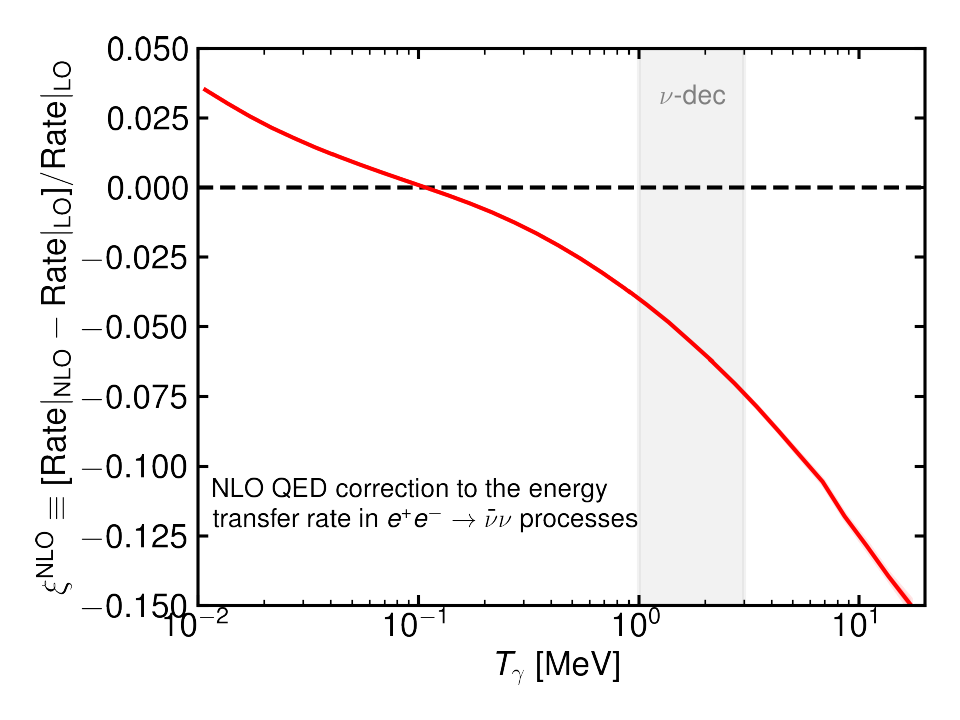}
\vspace{-0.4cm}
\caption{
The relative difference in the $e^+e^-\to \bar{\nu}\nu$ energy transfer rate between the NLO QED and Born scenarios as a function of temperature as calculated in this study. In grey, we highlight the approximate epoch at which neutrinos decouple in the early Universe.
}
\label{fig:NLOrate}
\end{figure}

The phase space of the NLO $e^+e^-\to \nu\bar{\nu}\gamma$ reaction rate is 15 dimensional, and this makes it unfeasible to solve for neutrino decoupling using the Liouville equation. Instead, we solve for neutrino decoupling using the integrated Boltzmann equations under the approximation that the distribution function of neutrinos is described by a Fermi-Dirac function with a temperature $T_\nu$ that is dynamically driven by neutrino interactions starting from energy conservation in the Universe. This approach has been shown to accurately reproduce all thermodynamic variables as compared with the actual solutions from the Liouville equation when using rates at LO~\cite{Escudero:2018mvt,EscuderoAbenza:2020cmq}. Given the agreement, in what follows, we use this method but including the NLO correction depicted in Fig.~\ref{fig:NLOrate}. The temperature evolution for the neutrino and electromagnetic temperatures reads

\begin{align}
 \frac{dT_{\nu_\alpha}}{dt} &= - H \, T_{\nu_\alpha} +   \frac{\delta \rho_{\nu_\alpha}}{\delta t}\bigg/\frac{\partial \rho_{\nu_\alpha}}{\partial T_{\nu_\alpha} }\, \label{eq:dTnudt}\,,\\
\frac{dT_{\gamma}}{dt}  &=- \frac{  4 H \rho_{\gamma} + 3 H \left( \rho_{e} + p_{e}\right) + 3 H T_\gamma  \frac{d P_{\rm int}}{d T_\gamma}+  \sum_\alpha\frac{\delta \rho_{\nu_\alpha}}{\delta t}}{ \frac{\partial \rho_{\gamma}}{\partial T_\gamma} + \frac{\partial \rho_e}{\partial T_\gamma} + T_\gamma \frac{d^2 P_{\rm int}}{d T_\gamma^2} } \, \, \label{eq:dTgammadt} .
\end{align}
where here $H$ is the Hubble rate, $\alpha = e\,,\mu\,,\tau$; $\rho_\gamma$ and $\rho_e$ are the energy densities in photons and $e^+e^-$, respectively, $p_e$ is the pressure of $e^+e^-$, $P_{\rm int}$ represents the NLO QED correction to the joint pressure of the electromagnetic fluid which we take from~\cite{Heckler:1994tv,Bennett:2019ewm}. Finally, the neutrino energy exchange rate can be written as
\begin{align}\label{eq:energyrates_nu_SM}
\left. \frac{\delta \rho_{\nu_\alpha}}{\delta t} \right|_{\rm SM}^{\rm FD} = \frac{G_F^2}{\pi^5} &[4\left(g_{\alpha L}^2+g_{\alpha R}^2\right) \, F(T_\gamma,T_{\nu_\alpha},\xi^{\rm NLO})  \nonumber \\
&+  \sum_{\beta} F(T_{\nu_\beta},T_{\nu_\alpha},1) ] \, ,
\end{align}
where $G_F$ is Fermi's constant, and as relevant for the energies of interest ($E\ll M_Z$)~\cite{Sarantakos:1982bp,pdg,Sirlin:2012mh,Erler:2013xha},
\begin{align}\label{eq:LowEnergyCouplings}
g_{eL} =0.727 \,,\qquad g_{eR} = 0.233 \,,\nonumber \\
g_{\mu L} = -0.273 \,,\qquad g_{\mu R} = 0.233 \,, 
\end{align} 
and where for $m_e = 0$ the $F$ function reads
\begin{align}\label{eq:G_FermiDirac}
\!\!\! F(T_1,T_2,\xi^{\rm NLO}) =&  32\, f_a^{\rm FD}   (T_1^9-T_2^9) \times (1+\xi^{\rm NLO})\\
+&56  f_s^{\rm FD}  T_1^4T_2^4  (T_1-T_2) \, \nonumber ,
\end{align} 
where the first term is related to annihilations and the second term is related to scatterings. Numerically, one finds $f_a^{\rm FD} = 0.884$ and $f_s^{\rm FD} = 0.829$. Here, $\xi^{\rm NLO}$ represents the NLO QED correction for the $e^+e^- \to \nu \bar{\nu}$ energy transfer rate, which we show in Fig.~\ref{fig:NLOrate}. Although the impact of the electron mass on these rates is small, we incorporate it by interpolating over the exact and numerically precomputed rates including $m_e$ as in~\cite{EscuderoAbenza:2020cmq}.  

We solve Eqs.~\eqref{eq:dTnudt} and~\eqref{eq:dTgammadt} including the NLO correction to the neutrino-electron energy transfer rate. We start the integration at a high temperature where neutrinos are strongly coupled, $T_\gamma = T_\nu = 20\,{\rm MeV}$\footnote{We have considered two cases, one where we track $T_{\nu_e}$ and $T_{\nu_{\mu,\tau}}$, separately, and one where we consider the neutrino fluid as coupled, $T_\nu = T_{\nu_e}=T_{\nu_{\mu,\tau}}$ and find the very same shift on $N_{\rm eff}^{\rm SM}$ from the NLO correction. The latter scenario can be seen as a way to artificially account for the effect of neutrino oscillations~\cite{Escudero:2018mvt,EscuderoAbenza:2020cmq}.}. We then solve the system until $T_\gamma \ll m_e$, and we explicitly check that the continuity equation is fulfilled at each integration step with high accuracy. As a result, we find that the inclusion of radiative QED corrections to $e^+e^- \to \nu \bar{\nu}$ rates implies a shift on $N_{\rm eff}^{\rm SM}$ of
\begin{align}
    N_{\rm eff}^{\rm NLO}- N_{\rm eff}^{\rm LO} = -0.0007\,.
\end{align}
This is the main result of our study. Combining it with the state-of-the-art calculations that include the effect of neutrino oscillations which yield a correction of the same size but opposite sign~\cite{Akita:2020szl,Froustey:2020mcq,Bennett:2020zkv}, one then finds
\begin{align}
    N_{\rm eff}^{\rm SM} = 3.0432(2)\,,
\end{align}
where the uncertainty is a combination of three factors of $0.0001$ added in quadrature. One comes from the numerical accuracy from the solution of the Liouville equation~\cite{Bennett:2020zkv}, another comes from current uncertainty on $\theta_{12}$~\cite{Bennett:2020zkv}, and in addition we have added another one because we have not accounted for radiative corrections in $e \nu \to e\nu$ reaction rates. The effect of scatterings on neutrino decoupling is a factor of $\sim 6$ smaller than that of annihilations, and expecting a change of at most $5\%$ in this rate due to NLO QED corrections makes up for this 0.0001 uncertainty. Finally, we note that while our calculation of the NLO correction does not include Pauli blocking in Eq.~\eqref{eq:Qloss}, the uncertainty generated by this should be at most $10^{-5}$ for $N_{\rm eff}$ and is therefore negligible. 

Finally, we would like to comment about the effect of other radiative corrections to the process of neutrino decoupling: 1) Electroweak radiative corrections -- it is important to properly consider the matching between the $SU(2)\times U(1)$ invariant theory at $E = M_W$ down to the four-Fermi theory describing neutrino-electron interactions at $E \sim 1\,{\rm MeV}$. The effect from virtual $W$ and $Z$ bosons is $\sim 1\!-2\%$ and has been calculated in the literature~\cite{Sarantakos:1982bp}. These effects are precisely taken into account at the one-loop level in the low-energy couplings describing the neutrino-electron interactions in our Eq.~\eqref{eq:LowEnergyCouplings}, see~\cite{Sarantakos:1982bp,pdg,Sirlin:2012mh,Erler:2013xha}. 2) Other processes induced at $\mathcal{O}(\alpha)$: at NLO in QED there are two other processes beyond the ones we have discussed so far: plasmon decays ($\gamma \to\bar{\nu}\nu $) and neutrino photoproduction ($e\gamma \to e\bar{\nu}\nu$). However, it is easy to see that plasmon decays are totally negligible in the early Universe at $T\sim {\rm MeV}$, see~\cite{Esposito:2003wv}, and that the rate of neutrino photoproduction at $T\sim {\rm MeV}$ is $\sim$ 0.1\% of that of $e^+e^-\to {\bar{\nu}}\nu$ annihilations and therefore is also negligible~\cite{Esposito:2003wv}. 

\section{Conclusions}\label{sec:conclusions}

Measurements of $N_{\rm eff}$ represent a key test for many scenarios beyond the Standard Model of Particle Physics and Cosmology and as such its prediction in the Standard Model is of remarkable importance. In this work, we have accounted for the last relevant ingredient needed to calculate $N_{\rm eff}$ with high accuracy: radiative QED corrections to the $e^+e^-\to \nu\bar{\nu}$ reaction rates. We have calculated this NLO correction as relevant for neutrino decoupling at $T\sim {\rm MeV}$. We have found that this correction is $\sim -5\%$ at $T\sim 2\,{\rm MeV}$, see Fig.~\ref{fig:NLOrate}, which suggests that $N_{\rm eff}^{\rm SM}$ should be slightly smaller than previously expected. 

By tracking the thermodynamic evolution of neutrinos, $e^+e^-$, and photons in the early Universe using the integrated Boltzmann equations including the NLO correction, we have found that indeed the effect of this QED correction is to reduce the number of relativistic neutrino species by $-0.0007$. Remarkably, in the Standard Model, this effect is as large in magnitude as that from neutrino oscillations. Combining this result together with calculations that account for neutrino oscillations and finite temperature corrections to the electromagnetic densities, we find that $N_{\rm eff}^{\rm SM} = 3.0432(2)\simeq 3.043$. This result is stable, and we believe that this NLO effect is the ultimate one needed to calculate $N_{\rm eff}^{\rm SM}$ with high accuracy since the current error bar on the Standard Model prediction is much smaller than any expected experimental sensitivity.

\vspace{-0.4cm}

\acknowledgements
We thank Mikko Laine and Jorge Martin Camalich for very useful discussions and suggestions over a draft version of the manuscript. M.E. would like to thank the Astroparticle Theory group at University of Naples {\it Federico II} and Istituto Nazionale di Fisica Nucleare (INFN), Naples, for the very kind hospitality extended to him when this study was initiated. The work of G.M and O.P. is partly supported by the Italian Ministero dell’Universit\`a e Ricerca (MUR), research grant number 2017W4HA7S “NAT-NET: Neutrino and Astroparticle Theory Network” under the program PRIN 2017 and the work of M.C., G.M, and O.P. by the INFN “Theoretical Astroparticle Physics” (TAsP) project. O.P. and G.M. acknowledge support by Ministero dell'Universita' e della
Ricerca (MUR), PRIN 2022 program (Grant PANTHEON 2022E2J4RK) Italy. O.P. and G.M. acknowledge support by Ministero dell'Universita' e della Ricerca (MUR), PRIN 2022 program (Grant PANTHEON 2022E2J4RK) Italy.

\newpage 

\bibliography{biblio}

\end{document}